# The Strong Coupling from Quarkonia*†

Aida X. El-Khadra

Department of Physics, Ohio State University, 174 W. 18th Ave, Columbus OH 43210

The status of determinations of $\alpha_s$ from quarkonia using lattice QCD is reviewed. We compare the results with those obtained from perturbative QCD.

## 1. INTRODUCTION

At present, the QCD coupling, $\alpha_s$, is determined from many different experiments, performed at energies ranging from a few to 90 GeV [1]. In most cases perturbation theory is used to extract $\alpha_s$ from the experimental information. Experimental and theoretical progress over the last few years has made these determinations increasingly precise. However, all determinations, including those based on lattice QCD, rely on phenomenologically estimated corrections and uncertainties, from nonperturbative effects. Table 1 (adapted from S. Bethke [1]) lists the caveats associated with each determination. These effects will eventually (or already do) limit the accuracy of the coupling constant determination. In the case of the lattice determination the limiting uncertainty comes from the (total or partial) omission of sea quarks in numerical simulations. We review the progress that has been made since the first determinations of $\alpha_s$ from the lattice [2,3].

The lattice determination of the strong coupling, $\alpha_s$, requires three ingredients:

1. The experimental input to the lattice determination is usually a mass or mass splitting, from which by comparison with the corresponding lattice quantity the scale, $a^{-1}$, is determined in physical units. For this purpose, one should identify quantities that are insensitive to lattice errors.

2. In the definition of a renormalized coupling, systematic uncertainties should be controllable, and at short distances, its (perturbative) relation to other conventional definitions calculable.

3. Calculations that properly include all sea quark effects do not yet exist. If we want to make contact with the "real world", these effects have to be estimated phenomenologically. One should therefore in "setting the scale" consider physical systems where the reliability of the phenomenological correction can be quantitatively estimated.

As has been pointed out some time ago by Lepage [2], quarkonia are the ideal systems for this program, since systematic errors can be analyzed using potential models.

Several groups presented results for quarkonium spectroscopy, which are reviewed by Davies [4]. The NRQCD collaboration [5,6] has results for the $b\bar{b}$ and $c\bar{c}$ spectra using a nonrelativistic QCD action at next to leading order. The leading order NRQCD action is used by the UK(NR)QCD collaboration [7] who presented results for the spin-averaged $b\bar{b}$ spectrum. The Fermilab group [3,8] has results for both the $b\bar{b}$ and $c\bar{c}$ spectra, using an $\mathcal{O}(a)$ improved Wilson action. The UKQCD collaboration [9], using a similar action, looked at the $c\bar{c}$ spectrum. The first calculation of the $c\bar{c}$ spectrum with dynamical ($n_f = 2$ staggered) fermions was presented by Onogi [10] for the KEK collaboration; they looked at the 1P-1S splitting using Wilson and staggered valence fermions.

Table 1
Comparison of various determinations of $\alpha_s$.

| process | observable | theory | caveats |
|---|---|---|---|
| $e^+e^-$ | had. event shapes, ... | pert. QCD (nlo, resummed) | hadronization |
|  | $R_{e^+e^-}$, $R_Z$ | pert. QCD (nnlo) |  |
|  | $R_\tau$ | pert. QCD (nnlo), analyticity | OPE, QCD sum rules |
|  | $\Upsilon$ decays | pert. QCD (nlo) | relativistic corrections |
| DIS | $F_3$ | pert. QCD (nlo) | higher twist |
|  | $F_2$ | pert. QCD (nlo) | higher twist, $g(x, Q^2)$ |
| $p\bar{p}$ | $p\bar{p} \to W+$ jets | pert. QCD (nlo) | large pert. correction |
| $b\bar{b}$, $c\bar{c}$ | spin averaged splittings | lattice QCD | sea quark effects |

## 2. SETTING THE SCALE

In setting the scale there are two sources of systematic error to be considered, apart, of course, from sea quark effects.

The leading lattice spacing errors of the relativistic Wilson action are spin dependent and therefore do not contribute to spin-averaged quantities. Likewise, for the non-relativistic action, spin-dependent interactions are added as higher dimensional operators; spin-averaged quantites do not depend on the additional couplings.

The scale, $a^{-1}$, depends in principle also on the other parameter which enters the calculation, the quark mass [4]. Table 2 shows that the experimentally observed mass dependence for spin-averaged splittings between the $c\bar{c}$ and $b\bar{b}$ systems is mild.

Figure 1 shows the mass dependence of the 1P-1S splitting in a quenched lattice calculation by the Fermilab group. Note that the splitting is about 10% smaller for $b\bar{b}$ than for $c\bar{c}$. Also shown in figure 1 is the (in-)dependence of the 1P-1S splitting on the improvement term. It is thus clear that spin-averaged splittings give the most accurate determination of the scale, $a^{-1}$. The NRQCD collaboration fits to the entire spin-averaged spectrum to determine $a^{-1}$ [4,5].

A summary of all the results for $a^{-1}$ is given in figure 2 as a function of the gauge coupling ($\beta \equiv 6/g_0^2$). The comparison of $a^{-1}$ from the $b\bar{b}$ and $c\bar{c}$ spectra shows a small scale dependence which is already apparent in figure 1. This ambiguity can only be resolved with the proper in-clusion of sea quark effects. In order to see how consistent the different results for $a^{-1}$ are with each other, we will apply in the following the same analysis in converting from $a^{-1}$ and $\alpha_{\text{lat}} \equiv 4\pi/g_0^2$ to $\alpha_{\overline{\text{MS}}}^{(4)}(5\text{ GeV})$.

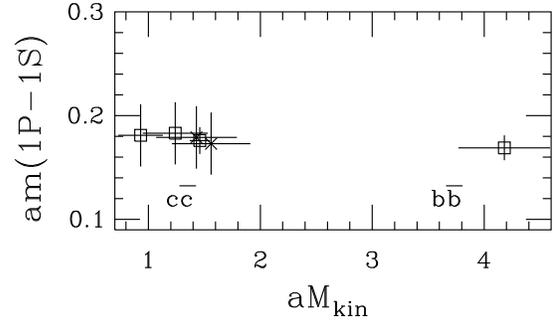

Figure 1. The 1P-1S splitting vs. kinetic mass on the $24^3 \times 48$ lattice at $\beta = 6.1$. $\times$: Wilson ($c = 0$) action; $\square$: Improved ($c = 1.4$) action.

Table 2
Spin-averaged splittings in the $\Upsilon$ and $J/\psi$ systems in comparison.

| quantity | $c\bar{c}$ (MeV) | $b\bar{b}$ (MeV) |
|---|---|---|
| $m(1P - 1S)$ | 456.8 | 452 |
| $m(2S - 1S)$ | 596 | 563 |
| $m(2P - 1P)$ | — | 359.7 |

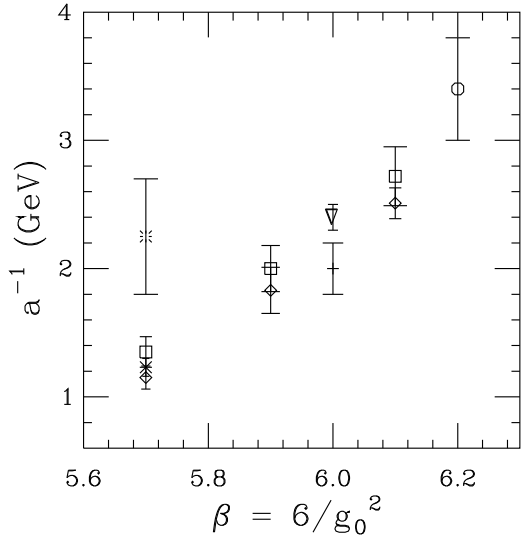

Figure 2. Summary of $a^{-1}$ vs. $\beta \equiv 6/g_0^2$ from the following groups: Fermilab [1] ($\square$: $b\bar{b}$, $\diamond$: $c\bar{c}$), NRQCD [2] ($\nabla$: $b\bar{b}$, $\times$: $cc$), KEK [3] ($*$), UK(NR)QCD [4] (o), and UKQCD [5] (+).

## 3. RENORMALIZED COUPLING

Within the framework of Lattice QCD the conversion from the bare to a renormalized coupling can, in principle, be made nonperturbatively. One can, for example, define a renormalized coupling from the nonperturbatively computed heavy quark potential (see the talk by Schilling [11]). In [12] a renormalized coupling is defined nonperturbatively through the Schrödinger functional; its running is computed using a finite size scaling technique. Of course, as always, the associated lattice spacing errors have to be carefully removed.

In ref. [13] it was shown that perturbation theory describes all short distance lattice quantities considered there up to a few percent provided a renormalized coupling is used. In ref. [2,3] the $\overline{\rm MS}$ coupling was determined from the bare lattice coupling using the mean field formula proposed in ref. [13]:

$$\alpha_{\overline{\rm MS}}^{-1}(\pi/a) = 4\pi \langle TrU_P \rangle / g_0^2 + 0.309 \ . \qquad (1)$$

An alternative is to define a renormalized coupling through short distance lattice quantities, like small Wilson loops or Creutz ratios. For example, the coupling defined from the the plaquette, $\alpha_P = -3 \ln \langle {\rm Tr} U_P \rangle / 4\pi$, can be expanded in $\alpha_V$ by [13]:

$$\alpha_P = \alpha_V(q)[1 - 1.19\alpha_V(q) + \mathcal{O}(\alpha_V^2)] \qquad (2)$$

or, equivalently

$$\alpha_V^{-1}(q) = \alpha_P^{-1} - 1.19 + \mathcal{O}(\alpha_V) \ , \qquad (3)$$

where $q = 3.41/a$ is chosen according to the prescription in ref. [13]. The $\overline{\rm MS}$ and $V$ couplings are related by:

$$\alpha_{\overline{\rm MS}}^{-1}(q) = \alpha_V^{-1}(q) + 0.822 \ . \qquad (4)$$

The size of higher order corrections associated with the above defined coupling constants can be tested by comparing perturbative predictions for short distance lattice quantites with nonperturbative results. It was shown in ref. [13] that, while the mean field coupling (1) accounts for the bulk of the corrections, $\alpha_V$ from the plaquette yields more accurate results. We will use both eqs. (2) and (3) for the analysis presented here, and average over the results.

In figure 3, the nonperturbative coupling from ref. [12] is compared to perturbative predictions for this coupling using eq. (3) and the 1-loop relations given in ref. [12]. Similarly, in ref. [5] the reliability of $\alpha_V$ using eq. (2) is estimated, from the spread observed in small Wilson loops and Creutz ratios up to size 3 compared to the plaquette. The next to next to leading order corrections for these quantities have been calculated numerically. We estimate a 5% uncertainty associated with perturbation theory.

Figure 4 shows the zero-flavor $\overline{\rm MS}$ coupling (using eq. (2)) as a function of $a^2$ for $b\bar{b}$ and $c\bar{c}$ for the different groups in comparison (with statistical errorbars only). The Fermilab results alone indicate that residual lattice spacing artifacts (higher than $\mathcal{O}(a)$) are too small to resolve with the statistical errors present. The scatter between results from the different groups, using either $b\bar{b}$ or $c\bar{c}$ states, is within the range of statistical errors. The small difference in $\alpha_s$ from $b\bar{b}$ as compared to

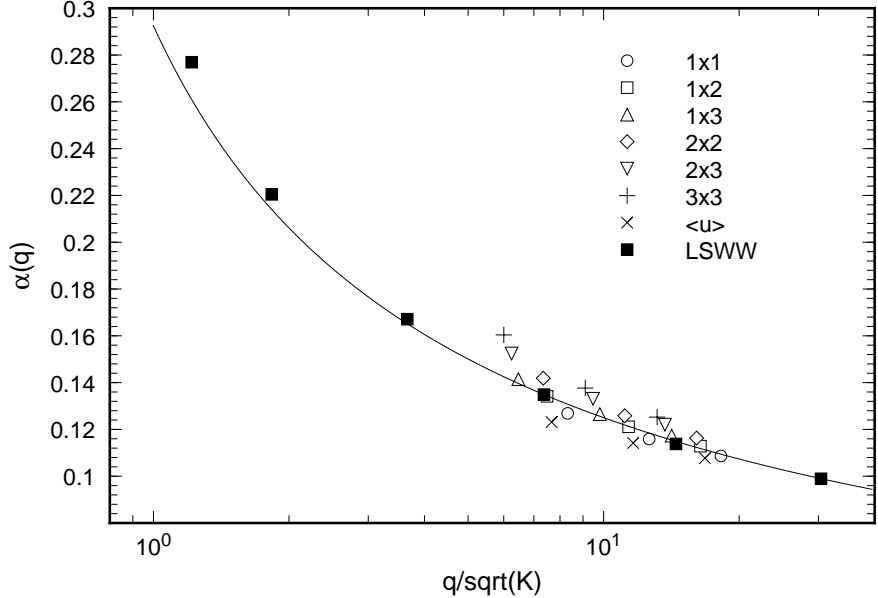

Figure 3. The nonperturbative coupling of ref. [5] in comparison with perturbative predictions.

$c\bar{c}$ that was observed in $a^{-1}$ is still present. Also, the 2 flavor coupling [10] is about 10% bigger than the quenched results.

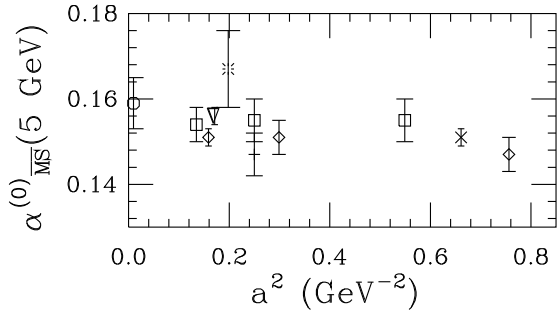

Figure 4. $\alpha^{(0)}_{\overline{MS}}(5\text{ GeV})$ vs. $a^2$. Fermilab ($\square$: $b\bar{b}$, $\diamond$: $c\bar{c}$), NRQCD ($\nabla$: $b\bar{b}$, $\times$: $cc$), KEK ($*$), UK(NR)QCD (o), and UKQCD (+).

In the next section we will show how, after proper inclusion of sea quark effects, everything falls into place.

## 4. SEA QUARK EFFECTS

This final correction can, at this point, not be included entirely from first principles. However, it can be estimated using potential model phenomenology and perturbation theory [3,14].

The physical quantity that is used to determine $a^{-1}$ sets the "matching scale", where by construction the coupling of the quenched lattice potential matches the coupling of the effective 3 flavor potential. The difference in the perturbative (2-loop) evolution of the 0 and 3,4 flavor couplings then causes the quenched coupling at short distances to be too small. This is illustrated in figure 5.

Different "matching scales" for $b\bar{b}$ and $c\bar{c}$ then lead to differences in the quenched coupling at short distances, as observed earlier. The matching scale is, in effect, the average gluon momentum transfer in the 'onium state in question. It can be estimated as the expectation value of the gluon momentum using potential model or lattice wave functions:

$$q^* = \langle q \rangle_{1S,1P,2S,\ldots} \qquad (5)$$

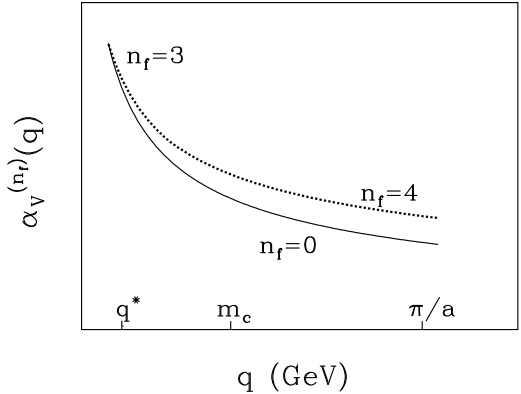

Figure 5. The 0 and 3,4 flavor running of $\alpha_V$ from a matching scale at $q = 700$ MeV to $q = 5$ GeV.

Lepage et al. [5] suggest to estimate $q^*$ as the expectation value of the potential. One could also use the scale at which the slopes of the $n_f = 0$ and 3 flavor potentials match. These different procedures lead to estimates for $q^*$ consistent with $350 - 650$ MeV for $c\bar{c}$ states and $600 - 1000$ for $b\bar{b}$ states. Note that the typical momenta in the p-wave states are more infrared than the momenta in s-wave states. The uncertainty in $q^*$ leads to an estimate of the uncertainty of the sea quark correction.

Before we go on to calculate the correction, a few words on the two-loop evolution of the coupling are in order. Lepage et al. [5] show that $\alpha_V$ runs with the 2-loop $\beta$-function down to momenta $\sim 700$ GeV (see figure 1 in ref. [5]). Something similar was also observed in ref. [12]. We thus conclude that 2-loop evolution is probably okay for $b\bar{b}$. We find a 20% correction, with a 5% uncertainty, to $\alpha_{\overline{MS}}$ at 5 GeV. However, close to the Landau pole, the perturbative coupling diverges and nonperturbative effects become important. Consequently, at the low end of the $c\bar{c}$ matching scales the perturbative correction becomes unstable and has to regulated with a potential model. We find a 26% correction with a 8% uncertainty.

A first principles test of the correction has now been performed for the first time by the KEK group, as shown in figure 2 of ref. [10]. The perturbative evolution of 0 and 2 flavor couplings to the charmonium scale gives consistent results, albeit with large errors.

Figure 6 shows the final result for $\alpha_{\overline{MS}}^{(5)}$, evolved to the Z mass scale, for the different cases (with statistical and systematic errors included). With the exception of the KEK result, the determination is totally dominated by the systematic errors from the quenched approximation and perturbation theory, leaving the other lattice errors neglegibly small in comparison. Thus, the quenched results are in much better agreement than the errorbars shown.

Table 3
Summary of uncertainties in the $\alpha_s$ determination at 5 GeV.

| uncertainty | $b\bar{b}$ (%) | $c\bar{c}$ (%) |
|---|---|---|
| stat., $a^2$ | $1 - 3$ | $1 - 3$ |
| perturbative | 5 | 5 |
| $n_f = 0 \to n_f = 3, 4$ | 5 | 8 |
| total | 8 | 10 |
| total at $m_Z$ | 5 | 8 |

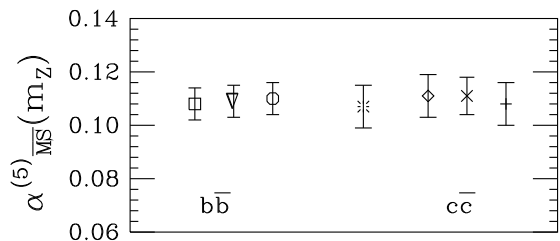

Figure 6. $\alpha_{\overline{MS}}^{(5)}(m_Z)$ in comparison; Fermilab ($\square$: $b\bar{b}$, $\diamond$ $c\bar{c}$), NRQCD ($\nabla$: $b\bar{b}$, $\times$: $c\bar{c}$), KEK ($*$), UK(NR)QCD (o), and UKQCD (+).

Table 3 lists all the uncertainties that enter the lattice determination of $\alpha_s$. We quote as our final result the value obtained from the $b\bar{b}$ spectrum:

$$\alpha_{\overline{MS}}^{(5)}(m_Z) = 0.108 \pm 0.006 \ . \tag{6}$$

The $c\bar{c}$ result, $\alpha_{\overline{MS}}^{(5)}(m_Z) = 0.110(8)$, is in good agreement with our central result. The difference between earlier results [2,3] and those presented here is almost entirely due to the change from eq. (1) to eqs. (2) and (3).

In figure 7 we show a comparison of results for $\alpha_s$ based on lattice and perturbative QCD. Since the analysis of ref. [5] is somewhat different from ours, the results labeled "NRQCD" in figure 7 are different from the values we quote but agree within errors. The quarkonium results are in good agreement with those from other experiments, with comparable uncertainties.

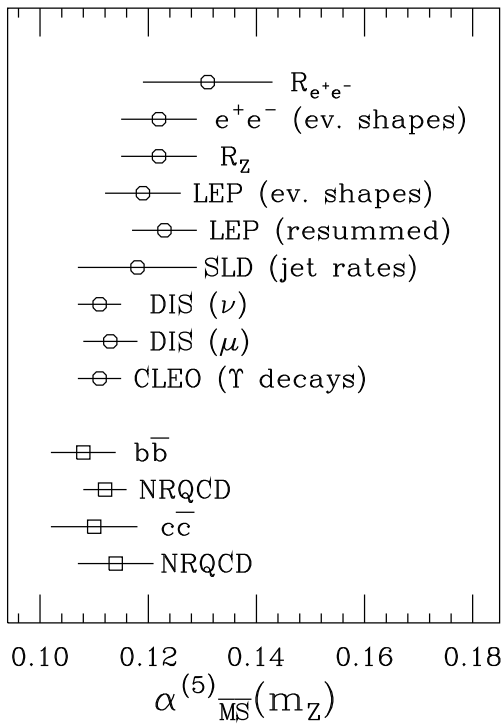

Figure 7. $\alpha^{(5)}_{\overline{\rm MS}}(m_Z)$ from lattice QCD in comparison with perturbative results. For recent references see ref. [1].

## 5. CONCLUSIONS AND OUTLOOK

An important goal of lattice QCD is the determination of Standard Model parameters from experimental input. Among the first results that include all systematic errors are determinations of $\alpha_s$ from quarkonium spectroscopy. This is an important achievement, even though the calculations are not entirely based on first principles.

The phenomenological correction that accounts for the systematic effects from the omitted sea quarks has been tested in two ways. The consistency of the results from $b\bar{b}$ and $c\bar{c}$ spectra is a phenomenological test of the correction. A first principles test with dynamical fermions was provided by the KEK collaboration [10], albeit with large errors.

As already mentioned in the introduction, phenomenological corrections are a necessary evil that enter most coupling constant determinations. However, in the long run, lattice QCD will give us a truly first principles determination of $\alpha_s$ with an accuracy that cannot be matched by any other method. At that point, the $\overline{\rm MS}$ standard should be replaced with a new one, a coupling that can be computed both in lattice QCD and perturbative QCD.

## 6. ACKNOWLEDGEMENTS

I would like to thank A. Kronfeld and P. Mackenzie, my colleagues of the NRQCD, UK(NR)QCD, UKQCD, and KEK collaborations, and especially P. Lepage, for discussions while preparing this talk.